**(1) Overview**

**BickGraphing: Web-Based Application for Visual Inspection of Audio Recordings**

**Paper Authors**
1. Seow, Kayley[1]
2. Arovas, Alexander[2]
3. Steinmetz, Grace[1]
4. Bick, Emily[1]

[1]Department of Entomology, University of Wisconsin-Madison
[2]Department of Biological Systems Engineering, University of Wisconsin-Madison

Author roles:
K. Seow: **Data Curation, Formal Analysis, Software, Validation, Visualization, Writing – Original Draft Preparation, Writing – Review & Editing**

A. Arovas: Conceptualization, **Data Curation, Methodology, Project Administration, Software, Supervision, Validation, Writing – Original Draft Preparation, Writing – Review & Editing**

G. Steinmetz: Data Curation, **Software, Validation, Visualization, Writing – Review & Editing**

E. Bick: Funding Acquisition, **Project Administration, Resources, Supervision, Validation, Writing – Original Draft Preparation, Writing – Review & Editing**

**Abstract**
BickGraphing is a browser-based research tool that enables visual inspection of acoustic recordings. The tool was built in support of visualizing crop-feeding pest sounds in support of the Insect Eavesdropper project; however, it is widely appliable to all audio visualizations in research. It allows multiple uploads of large .wav files, computes waveforms and spectrograms locally, and supports interactive exploration of audio events in time and frequency. The application is implemented as a SvelteKit and TypeScript web app with a client-side signal-processing pipeline using WebAssembly-compiled FFmpeg and custom FFT utilities. The software is released on [an open Git repository](https://github.com/bicklabuw/BickGraphing) (https://github.com/bicklabuw/BickGraphing) and archived under a standard MIT license and can be reused for rapid visual quality checks of .wav recordings in insect bioacoustics and related fields. BickGraphing has the potential to be a local, easy-to-use coding-free visualization platform for audio data in research.

**Introduction**
Audio data is of increasing importance to academic researchers. This is in part enabled by cost effective hardware, as well as easy to implement large data algorithms. Prior to deploying these algorithms, there is a critical step of exploratory data analysis. Sound lends itself to visualizing during this exploratory step.

Exploratory data analysis of audio includes visualizations of waveforms and spectrograms. Waveforms are visualizations of the recoding's time on the x-axis by amplitude on the y-axis. Spectrograms are 3D graphs that include time on the x-axis, frequency on the y-axis, and power serving as a colour coded intensity layer. These two basic visualizations are the most used method of exploring audio data (Ghadiri et al., 2017).

There are many methods to visualize waveforms and spectrograms, which include coding packages, paid software, and freeware. These programs were primarily designed for music, rather than research; this misalignment of use cases usually results in limited functionality when researchers try to visualize many files, and difficulty around the detailed inspections of each research audio file. Moreover, these approaches are limited by coding knowledge, funding associated with paid software, and the need to download software. Additionally, significant computational power is needed for all these methods. This severely limits potential field or in situ applications of exploratory analysis for acoustic sensors, such as checking if an acoustic sensor is functioning as expected in the field.

In response to this need, authors developed and launched BickGraphing, a web-based audio visualisation platform. BickGraphing has the potential to be a local, easy-to-use coding-free option for visualizing audio data in research. This platform fills this gap by providing a browser-based application that allows users to upload .wav files locally and interactively explore waveforms and spectrograms. In support of accessibility, offline use cases are supported with BickGraphing, via preloading or downloading. Here, we describe the software architecture, its availability, and its potential for reuse across audio research studies that require rapid spectrogram and waveform inspection.

**Implementation and architecture**

BickGraphing comprises four subsystems orchestrated by the central [graph.svelte](graph.svelte) component (Fig. 1): file input, signal processing, graphic visualization, file output.

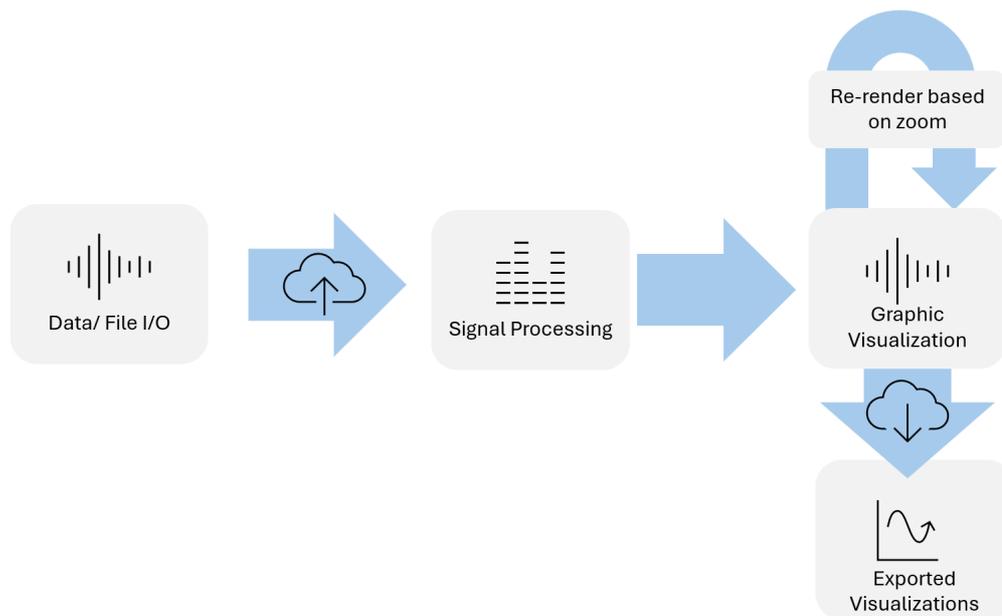

Fig. 1

### BickGraphing implementation

BickGraphing has been developed in SvelteKit using TypeScript because of its fast performance on web platforms, compile-time type checking, and ease of deployment as static files. It is implemented as a three-page web application, built and bundled with Vite for sub-second hot reloads and optimized production builds. It is styled with Tailwind CSS, with supporting plugins for forms and typography, for responsive, mobile-first layouts that work across desktop, tablet, and field-deployed tablets.

Having a 100% client-side execution eliminates server costs, and enables fast, efficient, and offline use in biological field settings.

### File I/O

BickGraphing's file I/O supports multi-file drag-and-drop workflows through the [FileSelector component](), which emits selected .wav files to the central [graph.svelte]() orchestrator. Duplicate detection and drag-reordering are handled, as well as file state management in [FileList](), enabling users to visualize multiple audio recordings simultaneously. Processed files populate the reactive state, triggering automatic AudioContext decoding and visualization generation without manual buttons("BaseAudioContext," 2024).

### Client-side signal processing

BickGraphing performs all audio decoding and analysis in the browser (Fig. 2). It uses a dual-decoding strategy that combines the native Web Audio API with FFmpeg to decode into Mono PCM (Pulse Code Modulation), raw digital audio samples represented as normalized floating-point values, for optimal performance and precision("BaseAudioContext - Web APIs | MDN," 2024; FFmpeg Project, 2024a).

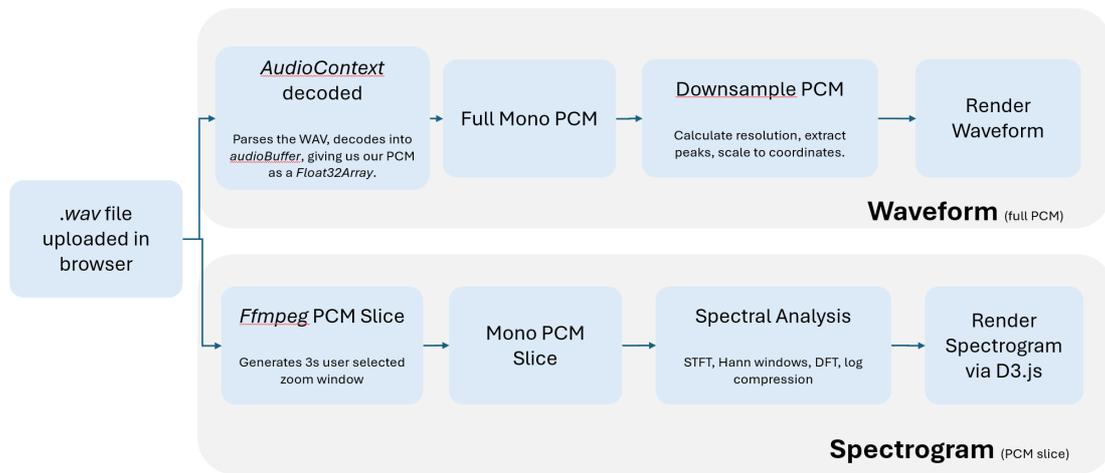

Fig. 2

### Waveform Generation: Native AudioContext Decoding

The Web Audio API's AudioContext.decodeAudioData() method provides instant full-file access to PCM audio data when a user uploads a .wav file, decoding it into an AudioBuffer where getChannelData(0) extracts the mono PCM stream as a Float32Array normalized between -1 to 1. Processing and downsampling the PCM results in the final rendered waveform("BaseAudioContext," 2024).

This API was chosen for having no external dependencies, asynchronous non-blocking execution on a dedicated audio thread, and hardware-accelerated decoding; the API delivers sub-second performance for >10-minute recordings across all major browsers, roughly 95% of the global share. This zero-configuration, standards-compliant implementation (Chrome 52+, Firefox 52+) makes it ideal for generating initial waveform overviews that enable rapid navigation through long environmental recordings ("BaseAudioContext," 2024).

### Spectrogram Generation: FFmpeg slicing & STFT Generation

FFmpeg extracts frame-accurate time windows from the original .wav file using the command -ss start_time -t duration into mono F32LE raw PCM. This delivers sub-millisecond slicing precision that reduces memory usage by 90%. This approach was chosen over AudioContext slicing due to its frame accuracy, multi-format robustness, and memory efficiency for STFT input. The FFmpeg enables precise spectrogram inspection of audio signals that would be impossible with browser PCM slicing alone (FFmpeg Project, 2024b). This on-demand approach powers detailed spectral analysis of specific time regions without loading entire recordings into memory, saving significant processing time.

### Spectral Analysis: Custom STFT Generation

Our custom Short-Time Fourier Transform (STFT) turns short audio chunks into frequency content using 2048-sample Hann windows (sliding every 1024 samples = 50% overlap) across the FFmpeg PCM slice. Each window converts to a frequency spectrum via Discrete Fourier Transform (DFT), stacking into a 2D grid showing time vs frequency (0–22kHz Nyquist range). Spectrogram processing takes ~10x longer than waveform generation due to the computational intensity of thousands of overlapping DFTs per second, window function multiplications, logarithmic compression, and 2D-to-3D colormap mapping across the full frequency spectrum. The log(1+magnitude) compression feeds D3.js turbo colormap rendering (with blue as low → red as high), clearly distinguishing sharp insect harmonics from broadband environmental noise (Fig. 3).

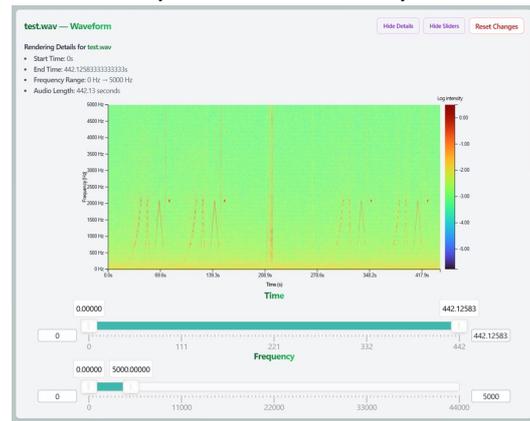

Fig. 3

While the waveform enables instant full-file navigation through AudioContext decoding, FFmpeg precision slicing powers detailed spectral zooms, balancing speed and accuracy. This dual-path architecture delivers sub-second file previews and frame-accurate spectrogram inspections entirely run on client-side. Running this signal-processing pipeline on the client reduces server requirements, enables offline field use, and avoids transferring large audio files over unreliable networks, a critical advantage for deployments in the field.

*User interface components and layout*

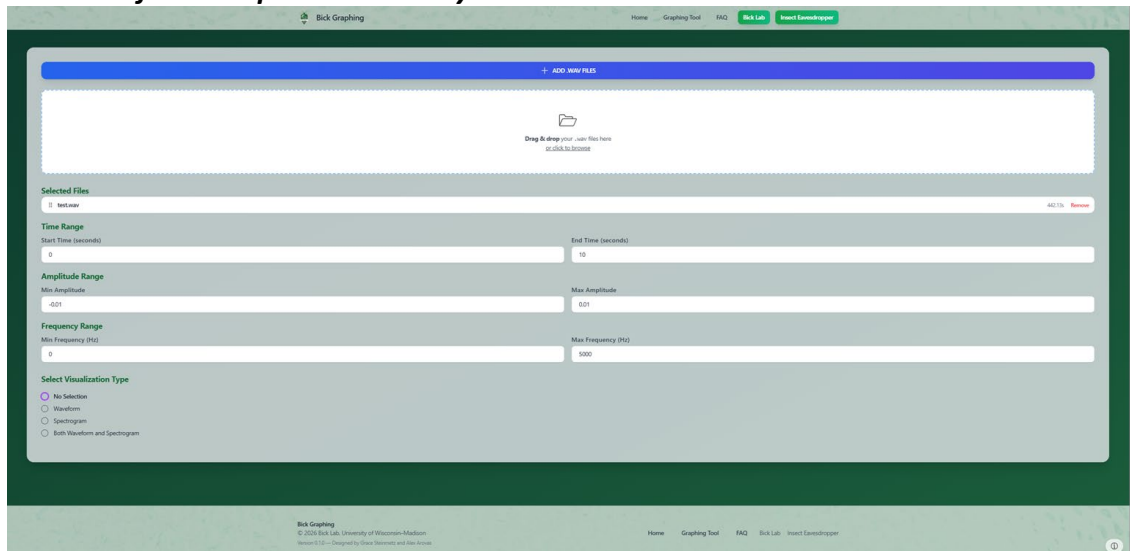

Fig. 4

BickGraphing's Svelte-based interface orchestrates signal processing and visualization through composable components managed by graph.svelte (Fig. 4). Key features:

File Management (Multi-File Workflow)

- [FileSelector](#) enables drag-drop uploads with duplicate filtering (processFiles()), while [FileList](#) supports drag-reordering (handleReorder) and per-file removal (removeFile()).
- Global controls [RangeInput](#) sync time/amplitude/frequency ranges across files (handleAllTimeChange).

View Control (Waveform/Spectrogram Toggle)
- [ViewSelector](#) toggles visualization modes via radio buttons (showWaveform/showSpectrogram), triggering generateVisualizations() recompute (handleVisChange).
- Miniwaveform thumbnails, [MiniWaveform](#), render in responsive 4-column grid for rapid multi-file scanning and selection(waveformDataMap).

Main Visualization (Interactive Graphs)
- [Waveform](#) delivers pan/zoom SVG plots from waveformDataMap (timeRangeMap/ampRangeMap), augmented by vertical [RangeSlider](#) sidebars for time/amplitude changes (handleTimeChange()), and [RangeInput](#) for precise selection.
- [Spectrogram](#) delivers the colour coded spectrogram,
- [ToggleButton](#) reveals rendering details; SVG download exports plots for easy sharing(downloadWaveform()).

BickGraphing prioritizes non-technical users through frictionless UX and instant feedback: drag-drop file processing eliminates buttons entirely ([FileSelector](#)), visual-first thumbnails and colour-coding replace numbers and terminals ([MiniWaveform](#)), progressive disclosure hides details by default ([ToggleButton](#)), and sub-second results deliver live slider updates in an offline-first workflow—taking farmers from file drop to actionable insights in seconds with no technical expertise required.

## Quality control

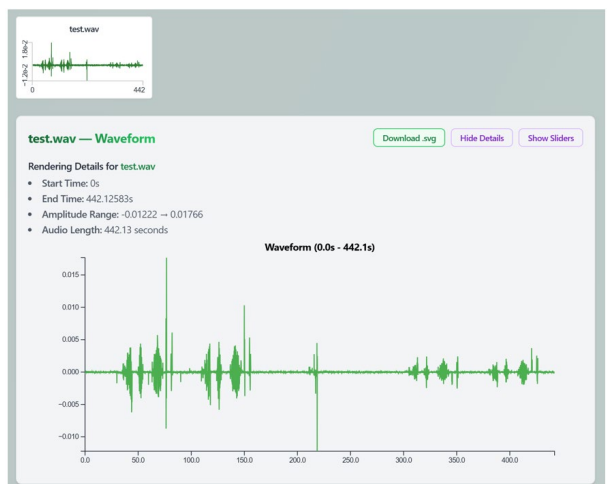

Fig. 5

BickGraphing emphasizes exploratory visual validation tailored to Insect Eavesdropper workflows rather than exhaustive automated tests. Core checks confirm .wav files decode correctly, waveform and spectrogram alignment in time and frequency, and interactive controls handle long recordings smoothly. Automated pipeline includes TypeScript type-checking, ESLint linting, and full builds via the npm run check, npm run lint, and npm run build commands. We have included a [test.wav file](test.wav file) for users to experiment with the software via loading, confirming that the waveform and spectrogram plots appear (Fig. 5).

**(2) Availability**

*Operating system*
Linux, Windows, and macOS systems running a modern WebAssembly-enabled browser; development and testing have focused on Ubuntu 22.04 and Windows 11.

*Programming language*
TypeScript (tested with TypeScript ≥ 5.x) using the Svelte and SvelteKit frameworks, built with Vite (≥ 6.x) and Node.js 20.x for development and tooling.

*Additional system requirements*
BickGraphing is designed to run efficiently in any modern web browser and does not require specialized hardware. To demonstrate its low computational footprint, the application was tested on an older-generation smartphone with limited processing capabilities and minimal available memory. Performance remained stable under typical use.

*Dependencies*
BickGraphing is implemented in TypeScript with SvelteKit and relies on the following main packages.
- Core framework and tooling
    - Svelte 5.0.0 or later
    - SvelteKit (@sveltejs/kit) 2.16.0 or later
    - Vite 6.0.0 or later
- Styling
    - Tailwind CSS 3.4.17 or later
    - @tailwindcss/forms 0.5.9 or later
    - @tailwindcss/typography 0.5.15 or later
    - @tailwindcss/vite 4.0.0 or later
- Audio processing
    - @ffmpeg/core 0.12.10 or later
    - @ffmpeg/ffmpeg 0.12.15 or later
    - @ffmpeg/util 0.12.2 or later
- Visualisation and interaction
    - D3 7.9.0 or later
    - nouislider 15.8.1 or later
    - svelte-dnd-action 0.9.61 or later
- Development and quality tools
    - TypeScript 5.0.0 or later

- ESLint 9.18.0 or later and typescript-eslint 8.20.0 or later
- Prettier 3.4.2 or later, with prettier-plugin-svelte and prettier-plugin-tailwindcss

*List of contributors*
1. Seow, Kayley[1]
2. Arovas, Alexander[2]
3. Steinmetz, Grace[1]
4. Bick, Emily[1]

[1]Department of Entomology, University of Wisconsin-Madison
[2]Department of Biological Systems Engineering, University of Wisconsin-Madison

*Software location:*
- **Code repository**
  - ***Name:*** BickGraphing
  - ***Identifier:*** https://github.com/bicklabuw/BickGraphing/tree/main
  - ***Licence:*** MIT License
  - ***Date published:*** 14/01/2026

*Language*
The repository, software, and supporting files are in English.

## (3) Reuse potential

BickGraphing was first designed for use with the Insect Eavesdropper acoustic sensor intended to monitor plant vibroscapes for insect activity (Bick, n.d.; Mehrotra et al., 2024; Schneider, 2024).To date, the Insect Eavesdropper has been evaluated on 27 insect species across multiple life stages on 17 cropping systems by more than 30 academic and government researchers. This platform enabled users of the Insect Eavesdropper to quickly validate whether the device was active and functioning as expected; it also provided immediate visualization of the plant vibroscapes, enabling immediate previews of results. BickGraphing's functionality enabled widespread generalizability for scientists working across systems in both laboratory and field settings.

Alongside quickly displaying waveform and spectrograms of audio files, this application's performance derives from intelligent subsampling implemented in generateVisualizations(), which produces multi-file waveform previews, while live sliders enable sub-second re-renders even for 30+ minute recordings. Completely offline and browser-based with minima memory footprint, BickGraphing is the choice for quick, portable audio data viewing.

The offline functionality enables BickGraphing to be useful in research settings with limited internet connections, such as for use in fieldwork. Additionally, the low systems requirements allow for widespread access to the platform.

Current pest monitoring technologies, such as image traps, pheromone lures, IoT sensors, and drones, typically demand connectivity, specialized equipment, or complex processing pipelines (Ahmed et al., 2024; Katherine et al., 2025; Reay-Jones et al., 2025). BickGraphing addresses this gap with a lightweight, economical solution for rapid vibroscape analysis.

BickGraphing can be reused by researchers who need fast visual inspection of .wav recordings across many domains, not only insect bioacoustics. In wildlife and conservation bioacoustics, scientists routinely scan spectrograms of long Passive-Acoustic Monitoring (PAM) recordings to locate bird, bat, and amphibian calls before or alongside automated detection, a workflow detailed in field guides. Birdsong and speech researchers similarly use waveform/spectrogram views to assess articulation, call structure and quality ("Capturing Bird Calls and Other Wildlife Sounds With Bioacoustics," 2020). Non-biological applications include in urban noise studies visually inspecting for traffic peaks or machinery noise, and engineering to check for machinery health via vibration profiling (Lan et al., 2020; Ma et al., 2022, 2022, "Urban Noise Pollution through Combined Analysis... | F1000Research," n.d.).

The modular Svelte and WASM architecture of BickGraphing allows it to be extended in several ways: developers can swap in different FFT parameters or window functions, add new visualizations (for example multi-channel or false-colour spectrograms), or integrate annotation and export tools that feed into downstream machine-learning pipelines.

The code for BickGraphing is publicly available and documented at Bick Lab's GitHub Repository (https://github.com/bicklabuw/BickGraphing/tree/main), maintained under the MIT License. The web platform removes coding barriers by simplifying the process of generating audio data visualizations. Support is not guaranteed for this code, but users are encouraged to contact the authors to discuss specific application possibilities, features, and issues at ebick@wisc.edu.

## Acknowledgements
Authors would like to thank Bick Lab Members Carmen Meyers, Dr. Mia Phillips, Raghav Jindal, and Praneet Popuri for their assistance in evaluating the BickGraphing Platform. We would also like to thank Dr. Kelsey Fisher (Connecticut Agricultural Experiment Station) and Dr. Dominic Reisig (North Carolina State University) for beta testing the program.

## Funding statement

If the software resulted from the Wisconsin Alumni Research Foundation Startup Accelerator Phase III grant and the Maldives Space Research Organization.

**Competing interests**
The authors declare that they have no competing interests.